# Field Trial of Alien Wavelengths on GARR Optical Network

Paolo Bolletta[*], Massimo Carboni[*†], and Gloria Vuagnin[*]
[*]GARR The Italian Research and Education Network, Rome, Italy
[†]INFN, National Laboratory of Frascati, Rome, Italy
Email: name.surname@garr.it

*Abstract*— At the Italian national level, GARR optical network is composed of two separate optical network domains. With the aim to integrate them we decided to use the alien wavelength technique. This is a hybrid solution based on the transmission and reception of optical signals, called alien wavelengths, that are transported on an infrastructure that is different from the one that generated them. We tested the technique first on a field trial, on two portions of our network of about 350 and 1200km long in order to measure the achievable performance. Based on the successful results we then implemented such technique on the production environment on about 3000km to deliver high-performance services. In this way we improved the overall efficiency of the Italian research and education network in a cost-effective way. This paper describes the overall activity, results and our experience in integrating the alien wavelengths in a production environment, with a special emphasis on deployment and operational issues.

## I. INTRODUCTION

GARR, the Italian research and education network, has recently started a process to further enhance its infrastructure and innovate the services provided, with the aim to match the ever evolving needs and requirements of its user community. Indeed, the general scenario characterized by an exponential growth of Internet traffic and a continuous strive for innovation, brought us to re-think our conventional network design and our common practices in devices engineering.

GARR responsibility to innovate while maintaining its production network encouraged us to adopt a strict method to introduce new technologies. This method consists in the following steps:

- solution design and feasibility study
- test in a dedicated environment
- field trial in production environment
- solution design optimisation
- deployment in the production network

These steps were followed in the activity described in the present paper. In section 2 we provide an overview of GARR user community and its network, section 3 describes the alien wavelength integration setup, sections 4 and 5 report on field trial and the related results, section 6 contains the details on how we deployed the alien wavelengths in the production network and finally in section 7 we draw some conclusions and provide some insights on the way ahead.

## II. GARR NETWORK OVERVIEW

### A. GARR Community

GARR is the Italian NREN (National Research and Education Network) connecting over 1000 sites all over Italy. Its user community is composed of universities, research institutes, research hospitals, cultural institutes, libraries, museums, and schools.

### B. GARR Optical Network

GARR optical network is based on two geographically separated infrastructures, which were deployed about four years apart and use different technologies. These two infrastructures led GARR to plan a performance study to evaluate the transmission of signals from the more recent technology over the older one.

The network infrastructure in Central and Northern Italy employs Huawei OptiX platform. The nodes include Reconfigurable Optical Add-Drop Multiplexers (ROADM) modules, add/drop boards able to support up to 80 channels in the C-band with a 50GHz grid and Optical Transport Network (OTN) switching matrices. Amplification is performed with Raman and Erbium-Doped Fibre Amplifiers (EDFA). Dispersion Compensating Modules (DCM) are inserted on the fibre lines in order to correct the chromatic dispersion. This optical network is optimised for Intensity Modulation with Direct Detection (IM-DD) and the channels are mainly 10Gbps with few 40Gbps. Client services are either at 1GEth or at 10GEth.

In Southern Italy, instead, the newer infrastructure is DCM-free and coherent technology is used for the transmission of signals. The network is equipped with Infinera DTN-X, a platform able to transmit 500Gbps super-channels. Each super-channel is built on ten optical carriers spaced at 200GHz in C-band on a 25GHz grid. Optical channels pairs can be enabled and managed with Quadrature Phase Shift Keying (QPSK) and/or Binary Phase Shift Keying (BPSK) modulation, thus allowing a flexible use of the available spectrum and an optimal balance between reach and capacity. Client services range from 10GEth to 100GEth.

*C. Alien Wave benefits for GARR*

In front of an ever increasing need for bandwidth by the user communities, 100GEth service capability is becoming a must for most NRENs. Moreover optical communications enhancements lead to more efficient spectrum use thanks to coherent transmissions which are taking advantage from digital signal processing. For these reasons, aiming at exploiting the coherent technology all over its network, GARR identified in the technique of alien wavelengths (AW) an effective and agile method to evolve its optical network. With the AW, the transponder light is sourced from a platform and is transported in the host optical domain regardless of the Dense Wavelength Division Multiplexing (DWDM) equipment vendor. Therefore the AW concept disaggregates the transponder elements from the optical DWDM system. This is beneficial as it allows the sharing of fibre resources among optical domains, thus allowing transponders from several vendors to be adopted and integrated over a photonic layer.

In the specific case of GARR optical infrastructure, the AW method made possible the integration of the two separated optical networks in one single domain, able to homogeneously deliver 100GEth services.

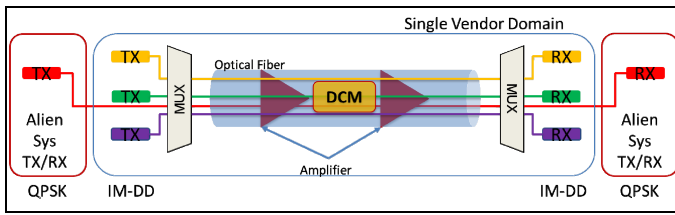

Fig. 1. Alien wavelength block diagram

Fig. 1 shows the main elements involved in the AW setup as functional blocks. Basically GARR deploys 100GEth services based on coherent signals over a fully compensated fibre infrastructure, by injecting signals at the service end-points and carrying the AW through the host network without regeneration and span redesign along the fibre lines. One of the main benefits of this approach is indeed the possibility to light up network services based on different technologies, without updating the geographical infrastructure and disrupting the original services, i.e. even keeping in place legacy network services. Moreover this approach requires new installations only at the service end-points, ensuring an effective and agile delivery phase.

### III. ALIEN WAVE SOLUTION SETUP

The integration between the two infrastructures: the one providing the transponders and the one providing the photonic layer is the most critical one and requires careful design and operational tuning. For GARR field trial, this was arranged as shown in Fig. 2.

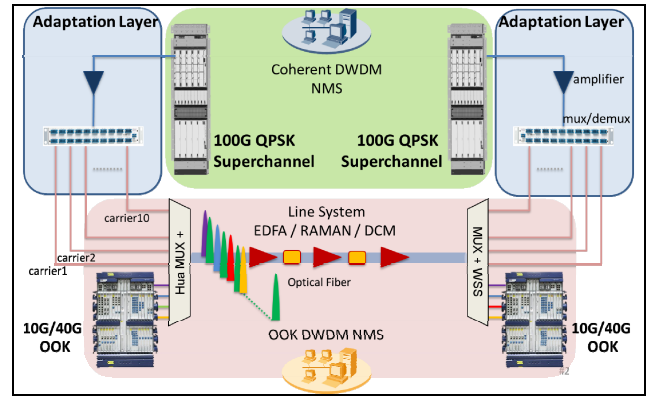

Fig. 2. Alien wavelength integration setup

The main building blocks of the system can be summarised as follows:

A. *Alien Wavelength domain:* Network Elements (NE) in charge of generating and operating AW signals, typically installed at service end-points. They are mainly composed of transponders mounted on: white-boxes, Data Centre Interconnect, OTN switches, etc.

B. *Native (host) domain:* NE that operate and control the photonic layer and the fibre infrastructure. The native domain composes the wide area network infrastructure for handling and managing the signals across the fibre footprint (Mux/Demux, Amplifiers, Wavelength Switch, ROADMs, ect.). If services are already deployed on the native infrastructure, the respective legacy elements and transponders will be part of this domain.

C. *Adaptation Layer:* Elements required to adapt the AW signals, in order to couple them with the host optical line system and to adjust the injected power at the proper levels.

### IV. THE FIELD TRIAL

The aim of the field trial performed by GARR was to test the coexistence of native Huawei 10Gbps IM-DD optical channels with coherent alien wavelengths of a higher bit rate. This field trial was carried out on a path of 345km and with an attenuation of 93dB (red path in Fig. 3). This path is equipped with DCMs modules and includes three points of amplification, one of which has a Raman amplifier. One of the two goals of this trial was to verify the ability of the existing GARR infrastructure to support at least one 100GEth client service provided by Infinera, only by adding the nodes at the end sites, regardless of the underlying infrastructure. The second goal was to determine the feasibility of a smooth migration from the less innovative DCM-based network towards the performance-enhancing coherent model.

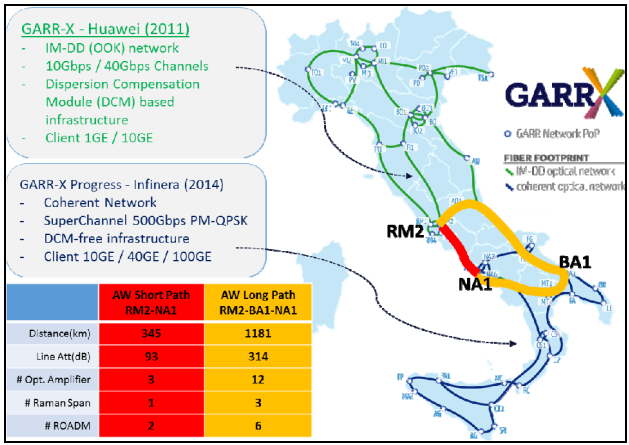

Fig. 3. Field test overview

A final test was performed on a longer path, which was 1181km long and with an attenuation of 314dB (yellow path in Fig. 3).

## V. FIELD TRIAL RESULTS

The field trial was carried out between March and April 2016 and the results were positive. A range of tests were performed, which differed in the use of guard bands between channels of the two different technologies; in the different coherent modulation schemes applied to the super-channel, and in the total reach achievable.

The test proved that the Infinera super-channel transported on Huawei amplification chains works and its performance is comparable to the one of a similar path totally equipped with a uniform Infinera infrastructure. Likewise the performance of Huawei channels sharing the same path with the coherent alien super-channel showed no negative impact. Also, the operations of managing the alien wavelengths as configuration of cross connections, alarms and performance monitoring on both network management systems were smooth and straightforward.

The final test performed on a longer path was also a success. The alien super-channel was configured with QPSK modulation and shared the path with 2 to 8 native lambdas over different segments of the path. Moreover, the throughput test performed on a 100GEth client service showed no errors.

Fig. 4 shows the use of the optical spectrum by means of an external appliance (Optical Spectrum Analyser – OSA). In particular it shows the coexistence, on the same portion of the spectrum, of native and alien wavelengths. Channel equalisation and power balance are performed in order to obtain homogeneous power levels among all the signals regardless of the modulation format and the spectrum occupancy.

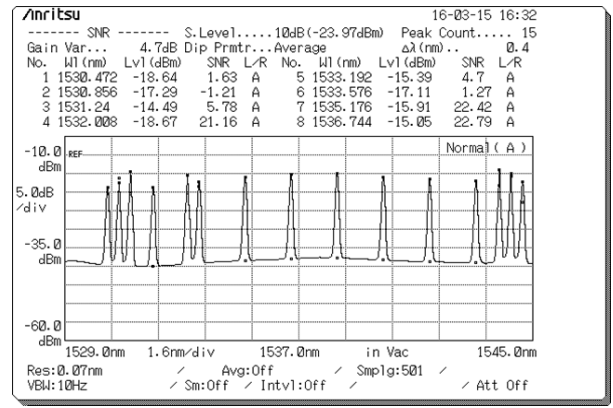

Fig. 4. Optical Spectrum occupancy

During field trial, the optical signal performance was evaluated at the receiver side, using the Qvalue metric and taking its average value on all the 10 super-channel carriers (below, average Qvalue). Furthermore, a color plan was designed without guard band between native and alien channels. This quite extreme scenario was considered in order to better evaluate the stability and effectiveness of this solution.

Fig. 5 shows the measurements of the average Qvalue of an AW super-channel as a function of the number of native signals sharing the same spectrum. These results pertain to the shortest path, where we studied the benchmark and the proof of concept for the solution.

The first point on Fig. 5 shows the measure taken when only the alien carriers were injected in the DWDM line, with no other signals present in the spectrum. The measure was taken after performing the line and channel equalisation, and can be considered as the best value for the alien performance in this specific set up. The next points of Fig. 5 were taken adding progressively one native channel (IM-DD), with (points 1-2) or without (points 3-5) guard band.

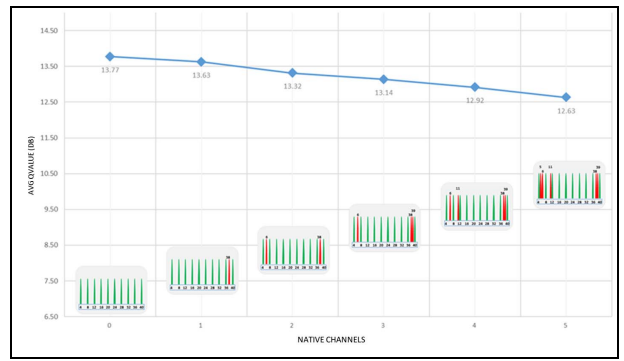

Fig. 5. Average Qvalue Alien+Native AWs

The result indicates that native channels affect AW average performance only marginally, even without guard band between native and alien channels. The penalties assessment during the tests were always within 0.5dB. It is important to remark that the Qvalue of a working signal should be higher than 6.5dB, however best practices suggest to work with a value above 8.5dB for a proper link design.

TABLE I. QPSK VS BPSK QVALUE

|  | Average Qvalue (dB) | |
| --- | --- | --- |
|  | *QPSK* | *BPSK* |
| AW | 13.77 | 16.37 |
| AW+2 native channels | 13.32 | 16.31 |
| AW+5 native channels | 12.63 | 16.15 |

TABLE I. summarises the measurements of average Qvalue achieved respectively by using either the QPSK or the BPSK modulation in three different cases: only AW; AW with native signal with guard band; AW with native signal, without guard band.

Fig. 6 shows a comparison of the measured alien coherent signal performance versus the link span. The figure shows a baseline of coherent super-channel performances in a native line system (in blue). This is a linear fit of five measures taken on similar boards on the GARR DWDM network based on Infinera and deployed in Southern Italy (blue diamants). We made the assumption that performances obtained on a well designed, engineered and deployed homogeneous system would be the best reference to compare the results of our tests.

The coherent super-channel performances were collected both on the shortest path and on the longest one, which has a distance comparable with the reach of the actual GARR production services.

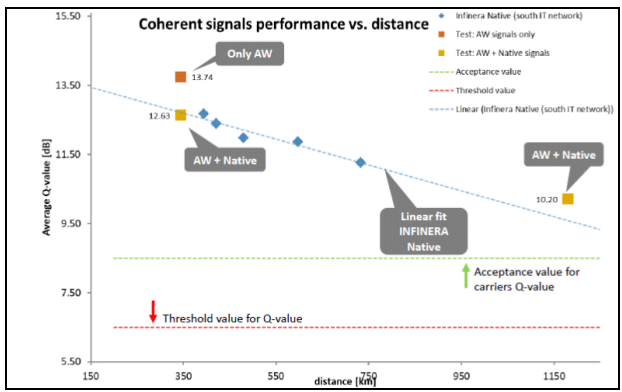

Fig. 6. Coherent signal performance vs. distance

The field trial measured points are:

1. Orange square: only AW transmission over approximately 350 km.

2. Yellow square: AW mixed with native signals over respectively over 350 km and 1200km.

This figure remarks that AW performance is in line with the reference, so we can conclude that, in the considered scenarios, the coupling of two different technologies with the AW method does not introduce relevant penalties with respect to traditional monolithic DWDM systems.

## VI. ALIEN WAVE IN PRODUCTION NETWORK

Thanks to these encouraging results, GARR decided to use the alien wavelength technique to provide 100GEth client services on the main backbone nodes of the transport infrastructure in the Northern and Central part of Italy. Indeed, the distances that interconnect the nodes located in Bari, Rome, Bologna, and Milan with a closed topology were compatible with the field-test results. The upgrade of GARR core optical backbone using the AW occurred during 2017, and had a positive outcome. From the perspective of the user community, the most important effect was an increased availability of bandwidth, as the coherent platform is far more spectrum efficient and enables 100GEth services in these nodes.

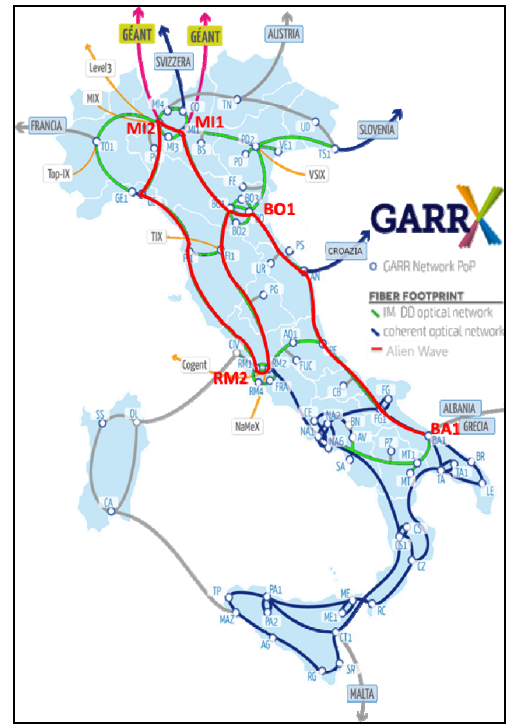

Fig. 7. GARR Alien wavelength production network deployment

From the network service perspective, the main outcome of the AW deployment is the full availability of 100GEth connectivity all over the core network nodes spanning from the South up to the North of the Country, reaching the main interconnection facilities (like upstream provider, cross border fibres and Internet Exchange Points). Furthermore merging and integrating DWDM networks offer different advantages in terms of IP/MPLS offload and decrease both CAPEX and OPEX for the service implementation and maintenance.

## VII. Conclusion and future plan

The Italian NREN GARR successfully performed a coherent alien wavelengths field trial over its compensated optical transmission infrastructure. This activity was aimed at evaluating the coexistence of 10Gbps IM-DD channels with high throughput coherent alien wavelengths. The goal of this trial was to verify a possible evolution path from GARR legacy DWDM system towards its new high-capacity coherent technology. The stability of the solution was demonstrated both on a 350km link and over a path that was about 1200km long. The results are positive and encourage GARR to consider the alien wavelength technique a valid solution to deliver high-capacity services all over its network. During 2017 GARR deployed the interconnection of the main 5 PoPs of its networks, by delivering 100GEth services over a 3000km optical network in less than 3 months. For GARR, This was only the first step towards the adoption of a disaggregated approach in optical networking, a vision that is currently under evaluation within its network infrastructure team.